\documentclass[journal]{IEEEtran}

\usepackage[utf8]{inputenc}

\usepackage{xcolor}
\usepackage{subfigure}
\usepackage{color}
\usepackage{graphicx}
\usepackage{balance}
\usepackage{amsmath}
\usepackage{amsfonts}
\usepackage{amssymb}
\usepackage{pifont}
\usepackage{times}
\usepackage{paralist}
\usepackage{textcomp}
\usepackage[hyphens]{url}
\usepackage[font={small}]{caption}
\usepackage{units}
\usepackage{amsmath,colortbl}
\usepackage{cite,comment}
\usepackage[nolist]{acronym}
\usepackage{soul}
\usepackage{array}
\usepackage{tabularx}
\usepackage{multirow}

\newcommand{\counting}[1]{#1} 

\begin{acronym}[ACRONYM]
\acro{3GPP}{the third generation partnership project}
\acro{5G}{fifth generation}
\acro{6G}{sixth generation}
\acro{AI}{artificial intelligence}
\acro{AoA}{angle-of-arrival}
\acro{AoD}{angle-of-departure}
\acro{BS}{base station}
\acro{D-MIMO}{distributed multiple-input multiple-output}
\acro{EM}{electromagnetic}
\acro{GDOP}{geometric dilution of precision}
\acro{GNSS}{global navigation satellite system}
\acro{IQ}{in-phase and quadrature}
\acro{JCS}{Joint Communication and Sensing}
\acro{JRC}{joint radar and communication}
\acro{JRC2LS}{joint radar communication, computation, localization, and sensing}
\acro{ICI}{inter-carrier interference}
\acro{IOO}{indoor open office}
\acro{IoT}{Internet of Things}
\acro{IRN}{infrastructure reference node}
\acro{KPI}{key performance indicator}
\acro{LoS}{line-of-sight}
\acro{MIMO}{multiple-input multiple-output}
\acro{mmWave}{millimeter-wave}
\acro{NLoS}{non-line-of-sight}
\acro{NR}{new radio}
\acro{OFDM}{orthogonal frequency-division multiplexing}
\acro{OTFS}{orthogonal time-frequency-space}
\acro{PRS}{positioning reference signal}
\acro{QoS}{Quality of Service}
\acro{RAN}{radio access network}
\acro{RAT}{radio access technology}
\acro{RedCap}{reduced capacity}
\acro{RIS}{reconfigurable intelligent surface}
\acro{RTK}{real-time kinematic}
\acro{RTT}{round-trip-time}
\acro{SLAM}{simultaneous localization and mapping}
\acro{SNR}{signal-to-noise ratio}
\acro{SOTA}{state of the art}
\acro{ToA}{time-of-arrival}
\acro{TDoA}{time-difference-of-arrival}
\acro{TR}{time-reversal}
\acro{TRP}{transmission and reception point}
\acro{TXRX}[TX/RX]{transmitter/receiver}
\acro{TX}{transmitter}
\acro{RX}{receiver}
\acro{UE}{user equipment}
\acro{multi-RTT}{multi-cell round-trip-time}
\acro{UL-TDOA}{uplink time-difference-of-arrival}
\acro{DL-TDOA}{downlink time-difference-of-arrival}
\acro{UMi}{3D-urban micro}
\acro{UMa}{3D-urban macro}
\acro{FR1}{frequency range 1}
\acro{FR2}{frequency range 2}

\end{acronym}
\begin{document}

\bstctlcite{IEEEexample:BSTcontrol}
\title{Positioning and Sensing in 6G:\\ \emph{Gaps, Challenges, and Opportunities}}

\author{Ali Behravan, Vijaya Yajnanarayana, Musa Furkan Keskin, Hui Chen, Deep Shrestha, Traian E. Abrudan,  \\Tommy Svensson, Kim Schindhelm, Andreas Wolfgang, Simon Lindberg, and Henk Wymeersch}

\twocolumn
\maketitle

\begin{abstract}
\counting{Among the key differentiators of 6G compared to 5G will be the increased emphasis on radio based positioning and sensing. These will be utilized not only for conventional location-aware services and for enhancing communication performance, but also to support new use case families with extreme performance requirements. This paper presents a unified vision from stakeholders across the value chain in terms of both opportunities and challenges for 6G positioning and sensing, as well as use cases, performance requirements, and gap analysis. Combined, this motivates the technical advances in 6G and guides system design. }

\end{abstract}


\IEEEpeerreviewmaketitle

\section{Introduction and Motivation}
\label{section:intro}


Large bandwidth and massive arrays employed in the emerging wireless communication networks along with network densification enable additional services, such as radio based positioning and sensing, which are beyond data transmission, with minimal cost by using the same infrastructure and spectrum. 
%
%
Positioning of {\em active} communication devices has become an integral part of the recent and ongoing standards,  such as in \ac{3GPP} and IEEE~ \cite{3GPP38857}. 
Position accuracy requirements have also been increasing from tens of meters, as mandated by regulatory agencies, to decimeter level for the future use cases such as indoor factories, unmanned aerial vehicles (UAVs), vehicle to everything (V2X), etc~\cite{hexax_d31}.
%
%
On the other hand, positioning of {\em passive} targets, i.e., sensing of objects that do not transmit (only reflect/scatter) radio signals has not yet been included in 3GPP standards. Radio based sensing covers a broad class of applications such as radar-like range and Doppler estimation, radio imaging, environmental monitoring and material identification. 
%
Hence, there is no single performance indicator and requirement that can be defined for a sensing service. As will be discussed, 
different use case families have different and new \acp{KPI} and require varying levels of sensing accuracy. 

In addition to supporting new use cases, another important motivation of integrated positioning and sensing in a mobile communication network is that such information about the environment can also be used to improve the communication performance. 
As an example, a digital twin of the environment that is created by sensing, can be used to aid communication functions such as radio resource management, beamforming, mobility management, minimization of driving test, etc. 

In a recent 6G localisation and sensing study conducted by the authors for the European Union Hexa-X project \cite{hexax_d31}, the potential of sensing with radio waves to enable new use cases and applications as well as improve communication aspects of the 6G systems are investigated. In this article, we highlight the key findings from these studies providing
a detailed gap analysis for positioning and sensing use cases in \ac{6G}, as well as envisioned 6G radio enablers and challenges, which serve to motivate continued research in this area.
\begin{figure}
    \centering
    \includegraphics[width=1\columnwidth]{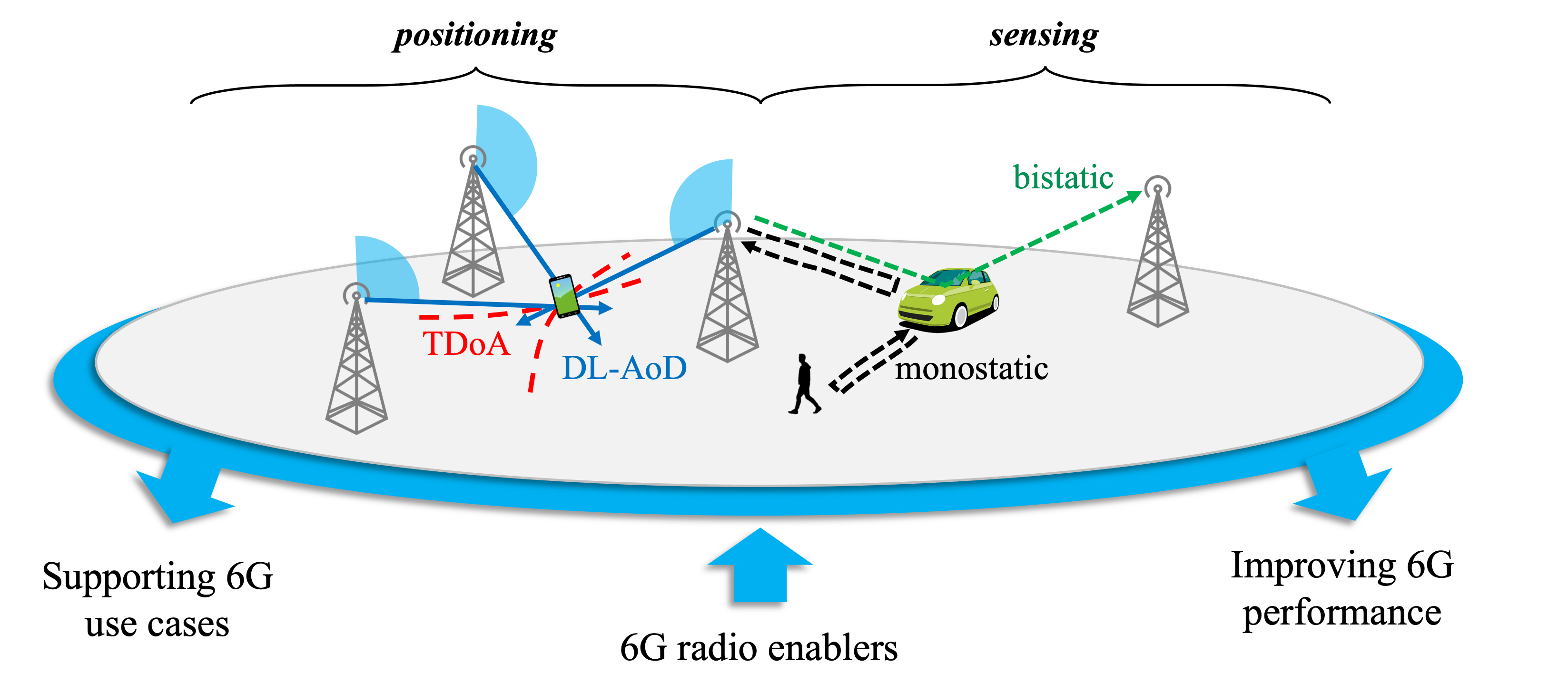}
    \caption{Positioning of a UE with several BSs (left) and sensing of an object (right). Positioning using TDoA from 3 BSs is shown in red dashed lines, constraining the UE to lie on the intersection of 2 hyperbola in 2D. Alternatively, DL-AoD (or UL-AoA) form lines which intersect at the UE location. Combinations of TDoA and AoD are also possible. Monostatic sensing by the BS (in DL) and the UE (in UL) 
    is shown in black, while 
    bistatic sensing is shown in green. }
    \label{fig:positioningandsensing}
\end{figure}

\section{Mobile Radio Positioning and Sensing} 
\label{section:overview}
In this section, we provide an overview of the fundamentals of mobile radio positioning and sensing.  

\subsection{Positioning and Sensing Fundamentals}


{\em Positioning} is the process of estimating the location (and in some cases, orientation and velocity) of a device from radio measurements such as received signal power, time-of-flight between transmitter and receiver, direction of the signal, or any combination of those, as illustrated in  Fig.~\ref{fig:positioningandsensing} (left part). The position estimation performance depends on the resolution and accuracy of the underlying measurements, number of \acp{BS} involved, and relative positions of the \acp{BS} with respect to the \ac{UE}. 
The cellular positioning reference signals employed in 5G include downlink positioning reference signal (DL-PRS) and uplink sounding reference signal (UL-SRS). Among the time and angle positioning methods, uplink/downlink time-difference-of-arrival (UL/DL-TDoA) use SRS/PRS, respectively, uplink angle-of-arrival (UL-AoA) uses SRS, and downlink angle-of-departure (DL-AoD) uses the beam index,  whereas multi round trip time (multi-RTT) relies on both PRS and SRS \cite{dwivedi2021positioning}.
%
In a typical 5G network, signaling information that is intrinsic to the network, such as serving cell, serving beam, timing-advance and reference signal received power (RSRP), can be exploited to estimate location information. Methods using such information are called \emph{enhanced cell ID (e-CID) methods}. The reference signals such as DL-PRS and UL-SRS can be employed to compute \ac{TDoA} between a pair of reference nodes (e.g., \acp{BS}), which can in turn yield a locus of points along a hyperbola in which \ac{UE} may be present. Thus, TDoA measurements from multiple pairs of reference nodes can triangulate to a precise \ac{UE} location. These are broadly termed as \emph{TDoA-based multilateration methods}. In addition, 5G mmWave operation enables large dimension MIMO, which can provide high angular resolution, thereby enabling precise angle information between the reference node and the UE. The angle measurements between the UE and multiple reference nodes can be exploited to arrive at the position of the UE. These are usually termed as \emph{angle-based positioning methods}. To avoid the need for tight synchronization in TDoA, 
\emph{round trip time (RTT) based methods} from multiple BSs have also been introduced \cite{3GPP38857}. 


{\em Sensing} is the process of detecting and tracking targets such as vehicles, obstacles and humans, etc., and estimating their relative range, velocity, size, shape, orientation or material properties. 
%
Depending on where the transmitter(s) and receiver(s) are placed, the sensing process can be classified as mono or bi/multi static sensing, as depicted in Fig.~\ref{fig:positioningandsensing} (right part). 
%
%
%
%
%
%
The 5G standard does not explicitly specify methods, signaling, and protocols for sensing; however, there have been some attempts in research to exploit 5G signals for sensing \cite{barneto2019full}. 
Considered use cases in these works include activity detection, presence detection, etc. It is envisaged that \ac{6G} may support new signaling, protocol and methods similar to positioning to support emerging new use cases for sensing.

\section{Use Cases and Requirements}
\label{section:usecases}
\begin{figure}
    \centering
    \includegraphics[width=0.7\columnwidth]{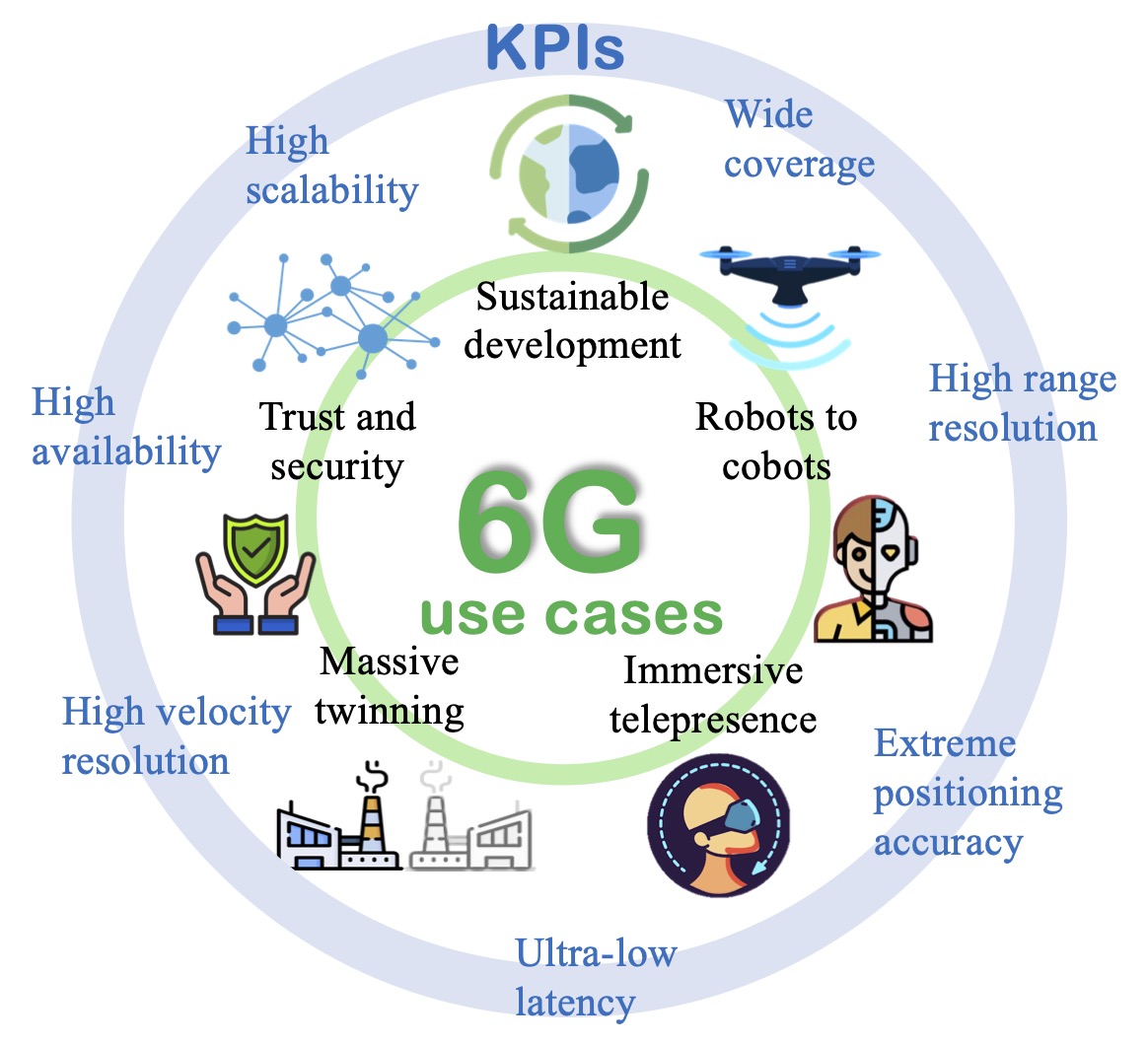}
    \caption{Use case families for 6G, covering a wide span of communication, positioning, and sensing requirements, together with the corresponding KPIs for positioning and sensing. }
    \label{fig:usecasefamilies}
\end{figure}

Potential use cases for the future wireless generation can be categorized into 5 use case families. An overview of these use case families together with the main positioning and sensing KPIs is presented in Fig.~\ref{fig:usecasefamilies} and elaborated below. Further details of the use cases together with some of the most commonly used metrics used for sensing and positioning are available in~\cite{hexax_d31} and references therein. Use cases are not fully orthogonal across the use case families. 

\paragraph*{Sustainable development} This use case family encompasses 6G use cases that address the sustainable development of society and, at the same time, reduce the environmental impact of different industries. One such use case is providing healthcare for all, regardless of geographical location. Remote healthcare is one such enabler which requires a rather fine location accuracy (e.g., drone deployment for medical sample collection). The required latency, in this case, is rather relaxed and a moderate availability is enough for a service guarantee.  
Other example use cases are remote sensing and monitoring of weather conditions, keeping track of biodiversity around the globe, and also asset tracking, all of which induce rather relaxed accuracy (meter-level) and latency (sec-level) requirements.

\paragraph*{Immersive telepresence} Sensing and positioning can improve immersive telepresence for enhanced interactions. This family of use cases includes gesture recognition for human-machine interactions and augmented reality (AR). Gesture recognition requires both rather fine range and also angular accuracy. Depending on the specific use case, AR may have different requirements. For example, while AR for providing context-aware services such as a shopping mall experience requires moderate location and angular accuracy, AR for placing an object in the real world requires cm-level location and tight orientation accuracies. 

\paragraph*{Local trust zones for humans and machines} Another use case family that can utilize sensing and positioning is information security and providing local trust zones for humans and machines. One part of this family includes use cases with very tight location and orientation, availability, and latency requirements,  such as telesurgery, localizing micro-robots within human body, as well as placement of medical equipment on the body. Another part of the family includes patient tracking and monitoring, sensor infrastructure web to support devices without sensing capability, cooperative positioning to support devices with little or no network coverage, and providing temporary local coverage when coverage from planned network infrastructure is not available, which can operate with more relaxed location accuracy, etc. 

\paragraph*{Massive twinning} Providing an efficient digital twin of objects and events in the digital domain can open up a host of new possibilities. 
One possible use case is in manufacturing, where configuring and using industrial tools can be done remotely. This type of use case requires location and range resolution in the order of centimeters. Also, this use case requires a good velocity resolution in the order of a fraction of a meter per second for moving objects. Another area where twinning can provide value is immersive smart cities, wherein one scenario, obtaining a real-time digital twin of the city can help optimize utilities. 
Another scenario in this area is traffic monitoring. Smart city use case requires meter-level to sub-meter level location accuracy, and rather relaxed latency and availability requirements. 
On the other hand, the digital twin of a smart building, where the location of each power switch, lamp, heater, etc., is important for effective interaction, requires more precise positioning, while tolerating longer delays and less availability. The most stringent requirement is the high scalability requirement. 

\paragraph*{Robots to cobots} This final use case family includes solutions that can enable collaborative robots (cobots), such as positioning of robots, obtaining a map of their environment, sensing objects, and fine positioning of vehicles around them. Some positioning use cases in this family such as localizing collaborative robots require very accurate position down to cm level, and also low latency requirements. 
Sensing applications, such as environmental mapping of robots, have tight sensing location accuracy requirements as well as velocity resolution, and angular resolution. 

\begin{figure}
    \centering
    \includegraphics[width=\columnwidth]{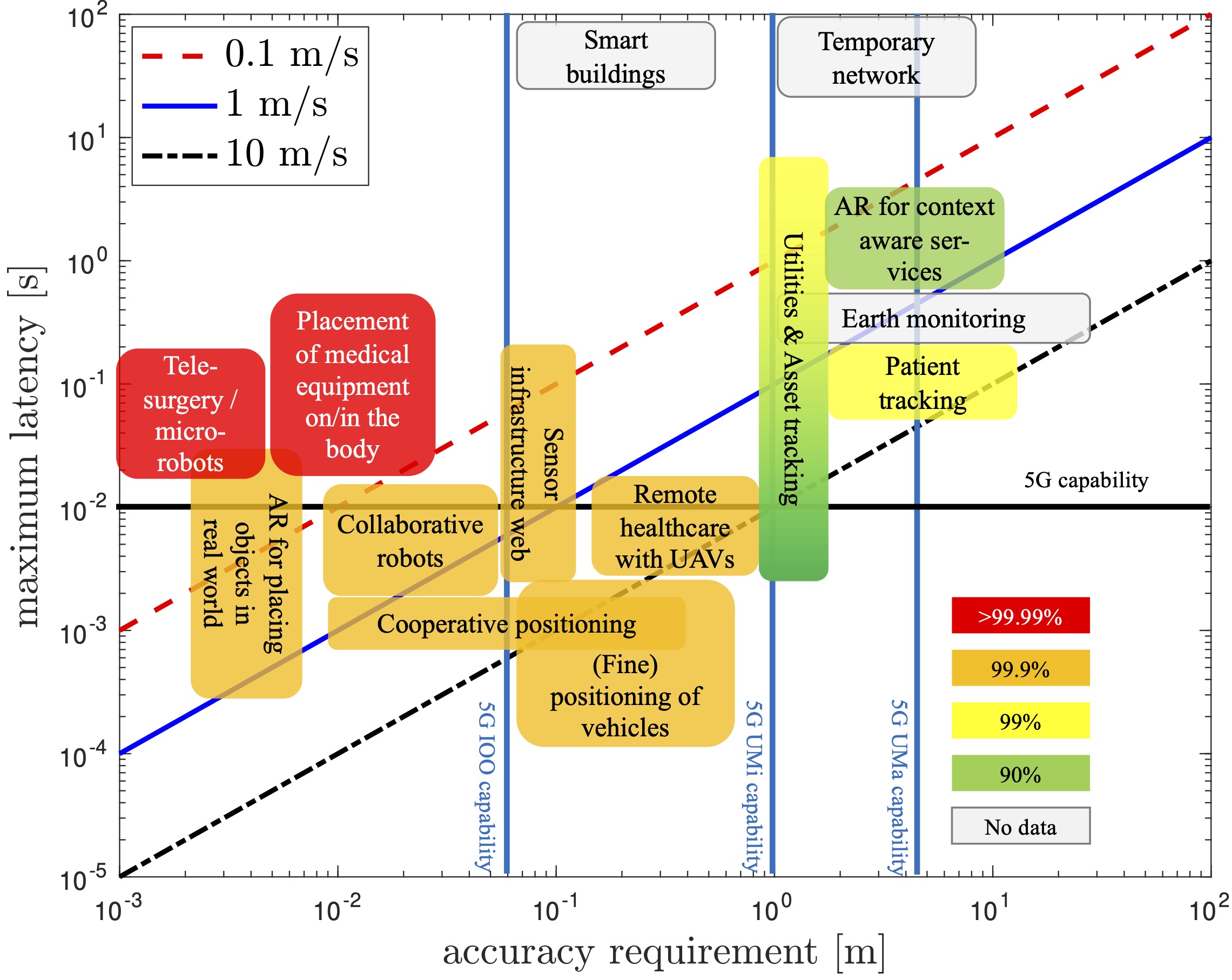}
    \caption{Accuracy, latency, and availability requirements for positioning use cases. The diagonal lines represent the maximum tolerable latency under the specified mobility, so that the accuracy requirement can still be met. Latency is determined based on the mobility, such that devices/objects move only 10\% of the target accuracy. }
    \label{fig:positioning_req}
\end{figure}

Fig.~\ref{fig:positioning_req} summarizes the requirements on accuracy, maximum latency, and availability for different use cases. 
The color-coding of use cases represent their availability requirements. These requirements have been derived based on the available literature (the full list is  available in~\cite{hexax_d31,siriwardhana2021survey}) and data provided by industry organizations. 

Table \ref{tab:sens_req} shows the requirements on location accuracy, range resolution, and velocity resolution for sensing use cases.


\definecolor{Gray}{gray}{0.9}
\begin{table}[hbt]
\caption{\label{tab:sens_req} Requirements for sensing use cases.}
\begin{tabular}{| m{3.2cm} | m{1.3cm}| m{1.3cm} | m{1.3cm}|}
\hline
\rowcolor{Gray}
\textbf{Use case} & \textbf{Location accuracy [m]} & \textbf{Range resolution [m]} & \textbf{Velocity resolution [m/s]} \\
\hline
Gesture recognition for human machine interface & 0.01 & 0.01 & 0.3 \\
\hline
Digital twins for manufacturing & 0.01 & 0.01 & 0.5 \\
\hline
Traffic monitoring & 0.5 & 0.5 & 0.5 \\
\hline
Robots to cobots environment mapping & 0.01 & 0.01 & 0.5 \\
\hline
Robots to cobots object sensing & \textless0.01 & \textless0.01 & 0.1 \\
\hline
\end{tabular}
\end{table}

\section{Gap Analysis}
\label{section:gap}

\includecomment{Should we classify use cases into different groups? Table II.}

\subsection{Positioning Gap Analysis}
In order to identify the gap between the requirements for the envisioned 6G use cases shown in Section~\ref{section:usecases} and the capabilities of existing technologies, we need to establish the achievable performance of state-of-the-art technologies in localization and sensing. For localization we use 3GPP Rel-16 as the baseline, and for sensing we consider the performance of the commercially available radar and lidar sensors. 

\subsubsection{Position Accuracy}
The positioning accuracy of different positioning methods, namely \ac{multi-RTT}, UL-\ac{TDoA} and DL-\ac{TDoA} in \ac{5G} \ac{NR} \ac{FR1} and \ac{FR2} is shown in Table~\ref{table:achievable_positioning_accuracy}.  Simulations are done for \ac{UMa}, \ac{UMi} and \ac{IOO} environments with 50ns synchronization error. It is seen that the achievable
accuracy for 90\% of the users with the best method can be
around $\unit[1]{m}$ for FR1 and $\unit[10]{cm}$ for FR2 in an IOO, and $\unit[3-4]{m}$ with FR1 in UMi and Uma.


\begin{table}[hbt]
\caption{\label{tab:pos_baseline} Achievable 5G positioning accuracies (in meters) for different methods and different deployment environments \cite{3GPP38857}. Cases with coverage limitation are shown as N/A.}
\begin{tabular}{| m{2cm} | m{.6cm} | m{1.3cm} | m{1.3cm} | m{1.3cm}|}
\hline
\rowcolor{Gray}
\multicolumn{2}{|c|}{\textbf{Method}}  & \textbf{UMa} & \textbf{UMi} & \textbf{IOO} \\
\hline
\multirow{2}{*}{DL-TDoA}  & FR1 & 4.37 m  &  3.48 m &  2.10 m  \\
\cline{2-5}
 & FR2 & N/A & 1.11 m & 0.17 m \\
\hline
\multirow{2}{*}{UL-TDoA}  & FR1 & 35.14 m  &  3.88 m &  2.19 m \\
\cline{2-5}
 & FR2 & N/A & N/A & 0.18 m\\
\hline
\multirow{2}{*}{Multi-RTT}  & FR1 & 30.29 m &  2.99 m &  1.11 m \\
\cline{2-5}
  & FR2 & N/A & N/A & 0.07 m \\
\hline
\end{tabular}
\label{table:achievable_positioning_accuracy}
\end{table}

5G capabilities for positioning accuracy are shown as vertical lines in Fig.~\ref{fig:positioning_req}. As it is seen from the figure, the required accuracy for the majority of use cases can not be met for the corresponding deployment scenario. For example the positioning accuracy that is required for remote healthcare which is supposed to be available for both indoor and outdoor scenarios can only be met by 5G indoor positioning capability.
Use cases demanding \textit{moderate accuracy} ($0.1-1$~m) can be supported by 5G positioning methods mainly in \ac{IOO}-like scenarios. 
Finally, the \textit{stringent positioning accuracy} ($<0.1$~m) requirements cannot be met in the considered scenarios.  


\subsubsection{Latency}
End-to-end positioning latency in 5G depends on the signaling delays between participating nodes and the positioning method employed. Latency evaluations for DL-TDOA, UL-TDOA, and multi-RTT are presented in ~\cite{hexax_d31}, which shows latencies on the order of $\unit[150-300]{ms}$ depending on the positioning method. Considering the most stringent latency for the  6G use cases, which is around $\unit[10]{ms}$, there is at least one order of magnitude gap between the state-of-the-art methods and envisioned 6G use cases.
5G capabilities for positioning latency are shown as diagonal lines in Fig.~\ref{fig:positioning_req} for different device speeds.



\subsubsection{Scalability}
5G positioning methods that are based on broadcast signals (e.g. PRS) can be used by multiple UEs or \acp{TRP} for positioning measurements, given that the nodes performing the measurements are in coverage of the broadcast signals. Hence, those methods are scalable and can be practically used to support the scalability requirements of the previously elaborated use cases. There is, however, an important limitation that will still be present in 5G. Accurate positioning requires radio resources (e.g., PRSs), which would thus occupy communication resources while locating. The higher the accuracy required, the larger the bandwidth of the channel that will be unavailable for communication. 
6G must solve this well-known problem by enabling simultaneous communication and positioning.

\subsubsection{Availability}
Future generations of cellular networks need to be designed to obtain seamless and pervasive connectivity in a variety of different contexts, matching stringent QoS requirements in outdoor and indoor scenarios with a cost-aware and resilient infrastructure~\cite{giordani2020toward}. The availability of the positioning system indicates that the position error (PE) is less than a threshold. High path loss is one of the major issues affecting the availability, which can be mitigated with the implementation of directional antennas and massive MIMO. Nevertheless, the angular coverage of antenna arrays (compared with omnidirectional antennas) will unavoidably be sacrificed. We expect new techniques such as reconfigurable intelligent surfaces (RIS),  distributed MIMO (D-MIMO), and scene-aware localization and sensing can meet the availability requirements. Considering rather limited availability of 5G, 
there is a significant gap between 5G capabilities and required availability for 6G use cases as shown by color mapping in Fig.~\ref{fig:positioning_req}, based on reported values in the technical literature (e.g., \cite{hexax_d31,siriwardhana2021survey}), when available. We observe 3 groups of requirements: use cases with not very stringent requirements, involving asset and people tracking and context-aware services; use cases involving robots and vehicles with strict availability requirements, and finally new medical use cases requiring extreme availability. 
Considering that the achievable availability of 5G is 90\% \cite{3GPP38857}, the availability for most of the use cases need to be improved.

\subsubsection{Orientation Accuracy}
The development of radio access technology-based positioning has been done considering regulatory and IIoT use cases, where UE orientation estimation is not a primary objective. For example, in regulatory use cases such as positioning of emergency call originating UE, knowing UE heading is not critical. This is partially due to the limited angular resolution provided by small antenna arrays. However, the next generation positioning use cases such as positioning for AR, positioning for collaborative robots, and local coverage for temporary usage greatly benefit from orientation estimation. In this regard, orientation accuracy should also be included in the primary objectives.

\subsubsection{Identified Gaps}
The position accuracy, latency, and availability gaps between 5G and 6G systems are summarized in Fig.~\ref{fig:positioning_req}.  
 To support the identified positioning use cases, future  \acp{RAN} must meet the requirements on the above-mentioned fundamental KPIs. 
The required positioning and orientation accuracy can be achieved only when the localization service is available, which is subject to coverage (indoor or outdoor), network deployment (\ac{GDOP} is one of the limiting factors), and synchronization between the network nodes. As a result, the next-generation positioning methods must, in principle, be able to address these issues and should offer more accurate position and orientation estimations by exploiting potential wide bandwidth and array size.
To achieve high scalability supporting much more devices in the future networks, it is of utmost importance to utilize the time (PRS allocation), frequency (bandwidth distribution), and spatial (beamforming and codebook optimization) resources with the highest efficiency. 
The latency budget evaluations show that the latency of 5G positioning methods depends on the employed method and can support many of the identified use cases. Nonetheless, the newly emerged use cases (e.g., enhanced remote health care, telesurgery, and collaborative robots) demanding quasi-real-time positioning mandate new technologies to obtain a much lower latency than what can be achieved by 5G systems.

\subsection{Sensing Gap Analysis}
In 5G there is so far no support for radio based sensing. To have a baseline against which to compare the expected performance requirements in 6G, we use radar and lidar. 
\subsubsection{Legacy solutions}
What can be achieved with radar and lidar technologies differs depending on the application and the environment. To have some numbers for comparison with the foreseen use cases in 6G, we have chosen Arbe radar, which can achieve a location accuracy and range resolution of around $\unit[0.5]{m}$ at a range of $\unit[300]{m}$ and a velocity resolution of $\unit[0.1]{m/s}$ with a latency of $\unit[33]{ms}$ \cite{hexax_d31}. The example we have chosen for lidar is the Hesai Pandar64, which can reach a position accuracy of $\unit[0.02]{m}$ at distances in the range $\unit[0.5-200]{m}$ with a latency of $\unit[50]{ms}$ and a surface with reflectivity of at least $10\%$ \cite{hexax_d31}. It does not provide any velocity estimate though. When comparing these values with the requirements for the different sensing use cases listed in Table~\ref{tab:sens_req}, a few things can be noted.

The lidar is good at positioning and is close to satisfying the location accuracy and range resolution demands of all the listed use cases, except for the object sensing in robots to cobots. Worth noticing though, is that the gesture recognition and the object sensing are performed at short ranges, which decreases the accuracy of the lidar. There is also a possible safety risk of having lasers pointing at a user from a very short distance as would be the case in gesture recognition. Another problem for the lidar is the lack of velocity measurements, meaning that it cannot fulfill the velocity resolution requirements for any of the use cases.

The radar is much better than the lidar in terms of velocity resolution and does actually fulfill the requirements for all the use cases. On the other hand, it lacks the precision in location accuracy and range resolution and only fulfills the requirements for the traffic monitoring use case. An advantage of the radar compared to lidar is that it is not as sensitive to the reflectivity of different materials in the surroundings. Possible health risks for users are also lower compared to the lidar.

\subsubsection{Identified Gaps}
\begin{figure}[!ht]
    \centering
    \includegraphics[width=0.95\columnwidth]{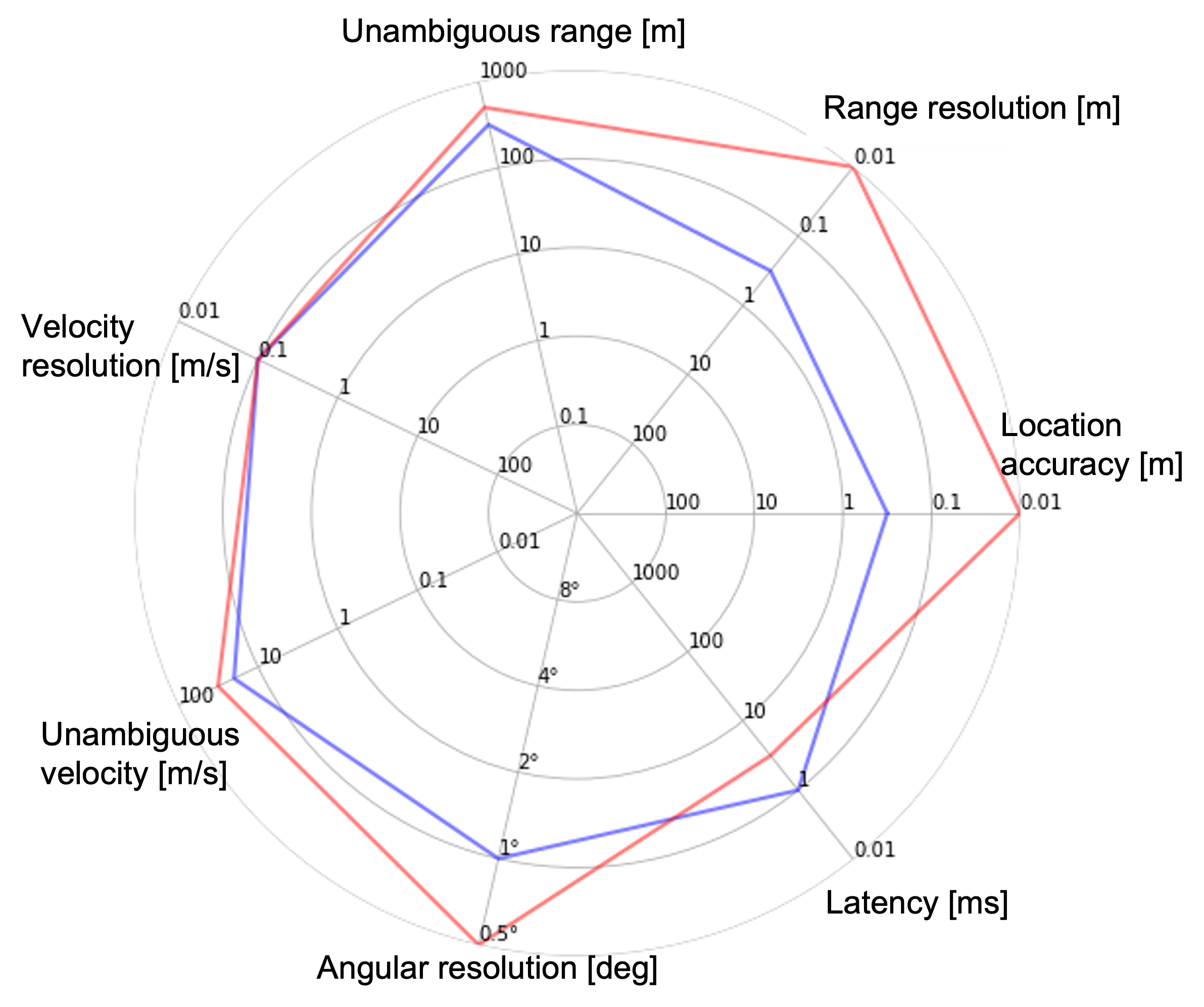}
    \caption{Sensing gap between legacy radar (in blue) and 6G (in red).}
    \label{fig:sens_gap}
\end{figure}

It is possible to satisfy most of the requirements listed in Table~\ref{tab:sens_req} using legacy radar and lidar. However, there is only one use case where it is possible to satisfy all requirements with only one technique, and that is road traffic monitoring. As was mentioned earlier, there exists a wide range of different radars and lidars, and in this case, we have only used two different examples for comparison. There are radars that perform better at positioning and lidars that give velocity information, but that is then at the cost of other performance metrics. The bottom line of this gap analysis is that there is no technique available today which satisfies all the requirements of the use cases.  To provide a fair evaluation of the gap between 6G requirements and the capabilities of legacy solutions, we consider radar as the baseline solution. Fig.~\ref{fig:sens_gap} shows the gap between the most stringent requirements for 6G use cases and the corresponding capabilities of legacy radar as mentioned earlier.

\section{Opportunities and Challenges for Positioning and Sensing in 6G}
\label{section:opportunities}
\subsection{Radio  Enablers}
\label{sec_enablers}

In order to close the gap between 5G capabilities and 6G requirements, various technical enablers, including high-frequency signals with large bandwidths and massive arrays, intelligent surfaces, and joint design of multi-functional hardware/waveforms have been considered. Below these are discussed in detail. 


\paragraph*{High-resolution Sensing with Large Bandwidths and Large Arrays}

The resolvability of multipath components in angle, range, and Doppler domains plays a crucial role in positioning and sensing. Higher carrier frequencies can accommodate larger bandwidth, resulting in superior range resolution. In addition, smaller wavelengths can bring significant antenna miniaturization, thus enabling the deployment of massive antenna arrays and leading to high angular resolution \cite{sippel2021exchanging}. This enables high-resolution sensing and mapping applications without being affected by ambient light and weather conditions, as opposed to visible light and infrared-based technologies \cite{rappaport2019wireless}.

\paragraph*{Reconfigurable Intelligent Surfaces}

As one of the key enablers in 6G, a \ac{RIS} can reflect an incoming electromagnetic wave towards a desired direction via programmable passive reflecting unit cells and a controller, which implies lower deployment and operational costs than a BS or a relay \cite{RIS_6G_2021}. 
Under line-of-sight (LoS) blockage conditions, RISs can create controllable non-line-of-sight (NLoS) links to improve coverage and communication quality. In positioning and sensing applications, RISs with known locations can boost accuracy by providing additional geometric measurements \cite{wymeersch2020radio}. Through tailor-made design of RIS phase shifts (i.e., passive beamforming), positioning and sensing performance can be enhanced significantly under a-priori knowledge of UE/target locations.

\paragraph*{Joint Hardware and Waveform Design}

Future positioning and sensing services shall rely on the ubiquitously available communication network and its hardware, thus avoiding the deployment of costly parallel infrastructure. Regarding joint waveform design, multi-carrier communication waveforms, such as \ac{OFDM}, are attractive for positioning and sensing thanks to wide availability and efficient implementation \cite{General_Multicarrier_Radar_TSP_2016}.
On the other hand, single-carrier waveforms can offer a better solution in terms of hardware efficiency due to low peak-to-average power ratio (PAPR), but may lead to higher side-lobe levels than \ac{OFDM}.
To investigate their resolution, accuracy and clutter rejection characteristics, waveforms (single- or multi-carrier) can be evaluated through range-Doppler ambiguity function.  
Due to inherent trade-offs, the joint waveform optimization for positioning, communications and sensing requires careful consideration of conflicting requirements, such as data rate, accuracy, and main-lobe width and side-lobe levels of the ambiguity function \cite{OFDM_DFRC_TSP_2021}.
Moreover, joint communications-sensing waveforms should be robust to hardware imperfections at high frequencies, necessitating simultaneous design of multi-functional transceiver hardware and waveforms.

\paragraph*{Algorithmic Developments}

With the high delay/angular resolution in 6G, two promising research threads for positioning/sensing algorithms arise in a complementary manner. \emph{Model-based} algorithms can exploit geometric optics in conjunction with optimization theory and statistical signal processing \cite{koivisto2017high}, while \emph{model-free} techniques rely on data-driven machine learning \cite{Wang2018RadioAnalytics}. Through their rigorous mathematical foundations and explainability, model-based methods seem attractive. Under severe hardware impairments and/or intractable mapping from measurements to position, data-driven approaches can become highly effective. In 6G scenarios, algorithms that can harness both data and domain knowledge will be key to achieving extreme positioning/sensing performance.

\begin{figure}
    \centering
    \includegraphics[width=\columnwidth]{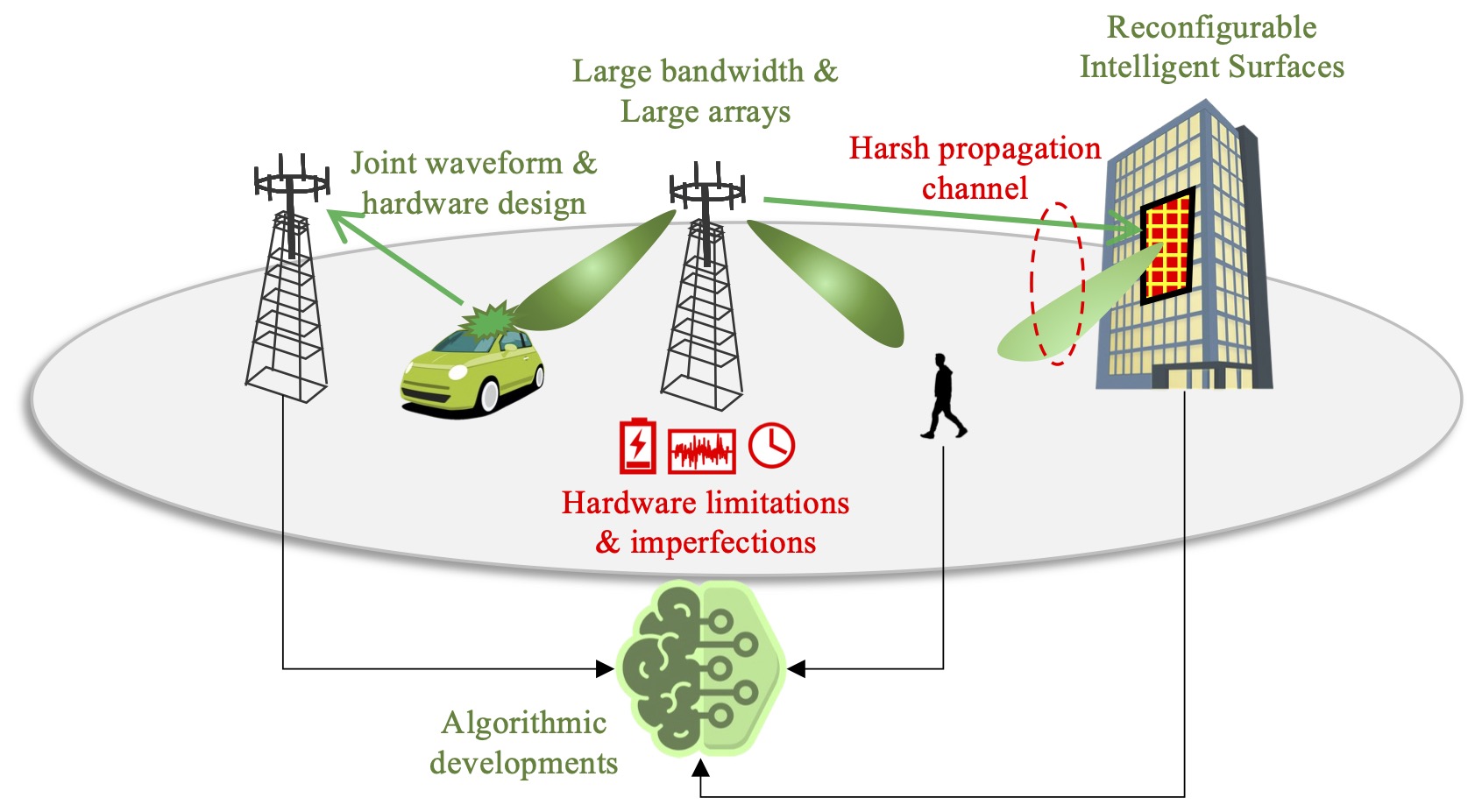}
    \caption{6G radio enablers that can help close the gaps in support of 6G use cases. Closing the gaps requires addressing two important  challenges.}
    \label{fig:6g_enablers_challenges}
\end{figure}

\subsection{Challenges} 


To fully harness the radio enablers for extreme performance, two fundamental challenges have been identified: hardware impairments and harsh channel conditions. 
In addition to those radio challenges, integrating highly accurate and low-latency positioning and introducing the totally new feature of integrated sensing pose further challenges and requirements, for example to the architecture and service offering concepts.

\paragraph*{Hardware Limitation and Impairments}

Hardware impairments bring more severe effects on positioning and sensing than on communication. Ranging accuracy degrades mainly due to timing errors, whereas angle estimation accuracy is degraded by antenna imperfections. Examples of hardware imperfections 
include phase noise, mutual coupling, nonlinear distortion, and frequency-selective impairments \cite{rikkinen2020thz}. 
While some of the hardware impairments can be compensated through  calibration (antenna mutual coupling and linear distortions), others, such as phase noise, must be compensated dynamically during operation.
%
In order to meet the low latency demands, a large bandwidth is required, but that is subject to hardware limitations. 
In addition, a large volume of data requires more storage and computational resources that can, in turn, limit the performance improved by these features. 

\paragraph*{Harsh Propagation Channel}


Depending on the relation between the wavelength and the size of objects, the radio channel can exhibit varying characteristics. 
At \emph{sub-6 GHz}, the channel has a very complex relationship to the environment and small movements lead to large power fluctuations due to small-scale fading. At \emph{mmW bands}, obstacle penetration is reduced, and reflection and scattering become more important phenomena. A sparser channel allows for higher multipath resolvability (characterized by fewer propagation clusters), and larger bandwidths and large antenna arrays enable accurate positioning. 
At \emph{100 GHz and above}, mainly multipath due to metallic objects will be visible, either in the form of moving incidence points or virtual anchors, or (groups of) smaller objects (e.g., pillars), behaving as static points. 
Additionally, the Doppler shift will greatly increase at the upper mmW frequency range. At such frequencies, the channel state changes faster requiring more frequent updates which demands better spatial consistency of the developed channel models so that consecutive channel impulse response samples are accumulated over time. 
Characterization of the angular, delay, and Doppler spreads due to extended objects, molecular absorption, link gains, and behavior under mobility are important challenges to be addressed, e.g., measured evidence and channel models.

\section{Outlook for Positioning and Sensing in 6G}
\label{section:conclusion}
In contrast to 5G, 6G will consider positioning and sensing as an integrated part of the system, with joint waveforms and hardware, as well as important cross-functional benefits. One of the important foreseen uses of 6G radio will be to support a wide variety of extremely challenging use cases, not only in terms of communication requirements, but also for positioning and sensing. The goal of this paper is to list a selection of these 6G use cases, determine their positioning and sensing requirements, and perform a gap analysis against the state-of-the-art. 
In terms of positioning, we reveal that in three key KPIs (accuracy, latency, availability), there is a significant gap between the 5G capability and the 6G requirements. In terms of sensing, we similarly found gaps in certain KPIs (accuracy and resolution) between state-of-the-art sensors and 6G sensing requirements.  To bridge this gap, this paper also presented a practical view of positioning for 6G radio, considering foreseen enablers (large bandwidths and arrays, RIS, algorithmic developments) and challenges (channel model mismatch and hardware impairments).

Next generation mobile networks 
will feature improved positioning performance,  and will enable sensing the environment. Through this, new services will arise using such information, which will have 
other implications. For instance, certain services will require  intermediate information with uncertainty and integrity guarantees, while other services will have security concerns that need to be carefully addressed, such as the access rights to different location services, location jamming, location falsification, as well as issues related to privacy and secrecy. 

Overcoming the challenges while harnessing the enablers in support of 6G use cases, will constitute a major research endeavor for the coming years. 
\counting{
\section*{Acknowledgments}
This work was supported, in part, by the European Commission through the H2020 project Hexa-X (Grant Agreement no. 101015956) and the MSCA-IF grant
888913 (OTFS-RADCOM).}

\bibliographystyle{IEEEtran}
\bibliography{IEEEabrv,reference}

\counting{
\vspace{5cm}


\begin{IEEEbiographynophoto}
	{Ali Behravan} is a master researcher at Ericsson research, Sweden, working in the area of wireless communication and sensing. 
  \end{IEEEbiographynophoto}
   \vspace{-1.2cm}
   \begin{IEEEbiographynophoto}
	{Vijaya Yajnanarayana} is a master researcher at Ericsson Research, India, working in the area of radio signal processing and artificial intelligence.  
  \end{IEEEbiographynophoto}
   \vspace{-1.2cm}
   \begin{IEEEbiographynophoto}
	{Musa Furkan Keskin} is a researcher and Marie Skłodowska-Curie Fellow (MSCA-IF) at Chalmers University of Technology, Sweden, working on joint radar-communications and mmWave positioning.
  \end{IEEEbiographynophoto}
   \vspace{-1.2cm}
   \begin{IEEEbiographynophoto}
	{Hui Chen} is a postdoctoral researcher at Chalmers University of Technology, Sweden, working on 5G/6G localization and sensing.
  \end{IEEEbiographynophoto}
   \vspace{-1.2cm}
   \begin{IEEEbiographynophoto}
	{Deep Shrestha} is a senior researcher at Ericsson research, Sweden, working on radio localization and sensing. 
  \end{IEEEbiographynophoto}
   \vspace{-1.2cm}
   \begin{IEEEbiographynophoto}
	{Traian E. Abrudan} is a senior researcher at Nokia Bell Labs, Finland, working on wireless communication and localization. 
  \end{IEEEbiographynophoto}
   \vspace{-1.2cm}
   \begin{IEEEbiographynophoto}
	{Tommy Svensson} is Professor at Chalmers University of Technology, Sweden, leading wireless systems research area with a focus on air interface and wireless backhaul/ fronthaul technologies for future wireless systems. 
  \end{IEEEbiographynophoto}
   \vspace{-1.2cm}
   \begin{IEEEbiographynophoto}
	{Kim Schindhelm} is a research scientist at Siemens' corporate research unit in Munich, Germany, working on localization. 
  \end{IEEEbiographynophoto}
   \vspace{-1.2cm}
   \begin{IEEEbiographynophoto}
	{Andreas Wolfgang} is CTO at Qamcom Research \& Technology, Sweden, working with communication and sensing. 
  \end{IEEEbiographynophoto}
   \vspace{-1.2cm}
   \begin{IEEEbiographynophoto}
	{Simon Lindberg} is a senior system developer at Qamcom Research \& Technology, Sweden, working with communication and sensing.
  \end{IEEEbiographynophoto}
   \vspace{-1.2cm}
\begin{IEEEbiographynophoto}
	{Henk Wymeersch} is Professor at Chalmers University of Technology, Sweden, working on radio localization and sensing. 
  \end{IEEEbiographynophoto}
}

\end{document}